%% file: main.tex
\documentclass[a4paper,onecolumn, 10pt]{article} 
\usepackage{amsmath,amssymb}
\usepackage[utf8]{inputenc}
\usepackage{authblk}

\usepackage{abstract}
\usepackage[a4paper, top=20mm, bottom=20mm]{geometry}
\usepackage[colorlinks,allcolors=black,bookmarks=false,hypertexnames=true]{hyperref} 

\usepackage[backend=biber, style=nature, sorting=none, doi=false,isbn=false,url=false]{biblatex}
\AtEveryBibitem{\clearfield{month}}
\AtEveryBibitem{\clearfield{day}}
\AtEveryBibitem{\clearfield{issue}}

\usepackage{textcomp}
\usepackage{setspace}

\usepackage{graphicx}
\usepackage{layout}
\usepackage[margin=10pt,font=small,labelfont=bf,format=plain]{caption} 
\renewcommand{\figurename}{Fig.}

\usepackage{color}
\usepackage{xspace}
\newcommand{\red}[1]{\textcolor{black} {#1}\xspace}

\newcommand{\beginsupplement}{%
        \setcounter{table}{0}
        \renewcommand{\thetable}{\arabic{table}}%
        \setcounter{figure}{0}
        \renewcommand{\thefigure}{S\arabic{figure}}%
        \renewcommand{\figurename}{Supplementary Fig.}
        \renewcommand{\tablename}{Supplementary Table}
     }

\title{Ultrafast Momentum-resolved Hot Electron Dynamics in the Two-dimensional Topological Insulator Bismuthene}

\author[1]{J. Maklar}
\author[2]{R. Stühler}
\author[1,3]{M. Dendzik}
\author[1]{T. Pincelli}
\author[1]{S. Dong}
\author[1,4]{S. Beaulieu}
\author[1]{A. Neef}
\author[5]{G. Li}
\author[1]{M. Wolf}
\author[1,6]{R. Ernstorfer}
\author[2]{R. Claessen}
\author[1]{L. Rettig}

\affil[1]{Fritz-Haber-Institut der Max-Planck-Gesellschaft, Faradayweg 4-6, D-14195 Berlin, Germany}
\affil[2]{Physikalisches Institut and Würzburg-Dresden Cluster of Excellence ct.qmat, University of Würzburg, D-97070 Würzburg, Germany}
\affil[3]{Current address: Department of Applied Physics, KTH Royal Institute of Technology, Hannes Alfvéns väg 12, 114 19 Stockholm, Sweden}
\affil[4]{Current address: Université de Bordeaux - CNRS - CEA, CELIA, UMR5107, F33405, Talence, France}
\affil[5]{School of Physical Science and Technology, ShanghaiTech University, Shanghai 200031, China}
\affil[6]{Institut für Optik und Atomare Physik, Technische Universität Berlin, Straße des 17. Juni 135, 10623 Berlin, Germany}

\date{\today}

\begin{document}

 \maketitle 
 
\begin{abstract}
Two-dimensional quantum spin Hall (QSH) insulators are a promising material class for spintronic applications based on topologically-protected spin currents in their edges. Yet, they have not lived up to their technological potential, as experimental realizations are scarce and limited to cryogenic temperatures. These constraints have also severely restricted characterization of their dynamical properties. Here, we report on the electron dynamics of the novel room-temperature QSH candidate bismuthene after photoexcitation using time- and angle-resolved photoemission spectroscopy. We map the transiently occupied conduction band and track the full relaxation pathway of hot photocarriers. Intriguingly, we observe photocarrier lifetimes much shorter than in \red{conventional} semiconductors. This is ascribed to the presence of topological in-gap states already established by local probes. Indeed, we find spectral signatures consistent with these earlier findings. Demonstration of the large band gap and the view into photoelectron dynamics mark a critical step toward optical control of QSH functionalities.
\end{abstract}

A promising platform for spintronic devices are two-dimensional (2D) topological insulators (TIs)~\cite{avsar2020colloquium}. Based on the quantum spin Hall (QSH) effect, 2D TIs feature an insulating band structure in their interior (here referred to as 2D bulk) surrounded by metallic states at their edges. These helical edge states (ESs) are characterized by spin-momentum locking, allowing for spin currents with opposite polarization for forward and backward-moving electrons. As they are topologically protected by time-reversal symmetry against single-particle backscattering, they also enable dissipationless transport~\cite{hasan2010colloquium, kane2005graphene}. However, any practical application based on helical ESs requires a large bulk band gap preventing interference with thermally excited bulk carriers at room temperature. Thus, in addition to a thorough characterization of the ES properties, also mapping out the bulk valence and conduction bands is critical. Yet, so far band structure investigations of 2D TIs are scarce~\cite{tang2017quantum, chen2018large, wu2016evidence} and an understanding of their dynamical properties and microscopic interactions\red{, imperative for controlling QSH functionalities,} is lacking.

A suitable approach to tackle these questions is time- and angle-resolved photoemission spectroscopy (trARPES), which has been pivotal for characterizing the electronic structure and fundamental interactions of 3D TIs, the 3D analogs of QSH insulators~\cite{wang2012,sobota2012ultrafast, sobota2014distinguishing, cacho2015momentum, neupane2015gigantic,  sumida2019magnetic, reimann2018subcycle}. This method grants direct access to the energy- and momentum-dependent electron dynamics after femtosecond (fs) optical excitation and to states that are not occupied in equilibrium. Thus, trARPES allows to map the transiently populated states above the Fermi level, which has been essential for a clear separation of semiconducting bulk and metallic topological in-gap states in 3D TIs. Gaining a similar understanding of the electronic structure and elementary scattering processes of 2D TIs is of strong interest both from a scientific and application perspective.

\begin{figure}[!th]
\centering
\includegraphics[]{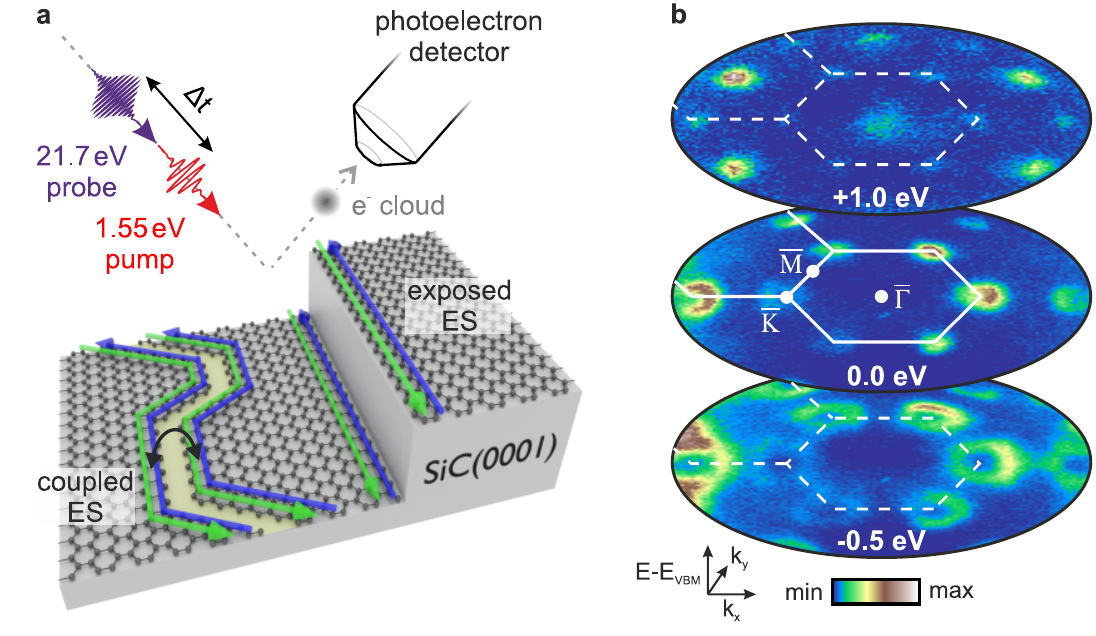}
\caption{Experimental scheme and photoelectron constant-energy contours. (a) Illustration of the trARPES experiment. An optical pump pulse excites the bismuthene sample, followed by an XUV pulse that probes the electronic distribution after a time delay $\Delta t$. The green and blue arrows represent the two spin channels of the coupled ESs at a domain boundary and of the exposed ESs at a substrate step edge. (b) Constant-energy contours with radius $k_\parallel \sim2$\AA$^{-1}$ of bismuthene after photoexcitation ($h\nu=1.55$\,eV, $\Delta t=-75\ldots+75$\,fs). Two exemplary BZs and high-symmetry points are indicated.}
\label{fig:overview}
\end{figure}

A novel platform to address this knowledge gap is the room temperature QSH candidate bismuthene, i.e., a monolayer of bismuth atoms arranged in a planar honeycomb geometry on a \red{semiconducting} silicon carbide SiC(0001) substrate~\cite{Reis2017BismutheneOA}. Spatially resolved scanning tunneling spectroscopy (STS) measurements have demonstrated a large band gap of $\sim0.8$\,eV in bulk areas, far greater than in any other QSH system~\cite{konig2007quantum, Knez2011_PRL, leubner2016strain, tang2017quantum, chen2018large, fei2017edge, wu2018observation, wu2016evidence, Sun2022Jan_ACSNano}, and conductive 1D states at exposed sample edges near substrate terrace steps~\cite{Reis2017BismutheneOA, stuhler2020tomonaga}, as illustrated in Fig.~\ref{fig:overview}a. Intriguingly, in topologically nontrivial materials, additional pairs of coupled ESs can arise within the bulk areas at extended 1D defects~\cite{lima2016topologically, pezo2021disorder}, recently observed within the 2D bulk of bismuthene along structure-induced domain boundaries~\cite{stuhler2020interacting}.  However, a demonstration of the elusive ESs using a momentum-resolved probe has proven challenging, as they constitute only a marginal fraction of the total surface area. Furthermore, \red{confirmation of the theoretically predicted large indirect bulk band gap of bismuthene and characterization of} microscopic carrier scattering processes \red{are} still missing\red{, as previous studies largely relied on momentum-integrating local probes}~\cite{Reis2017BismutheneOA, stuhler2020tomonaga, stuhler2020interacting}.

Here, we investigate the ultrafast electron dynamics of photoexcited bismuthene at room temperature using trARPES, as illustrated in Fig.~\ref{fig:overview}a, allowing us to access the microscopic scattering channels from the dynamics of the nonequilibrium state prepared by the optical excitation. Combining a hemispherical analyzer and a time-of-flight momentum microscope for photoelectron detection~\cite{puppin2019, maklar2020}, we map the transiently populated conduction band structure and confirm the existence of a wide indirect bulk band gap of $\sim0.82$\,eV. In addition, we identify faint gap-filling spectral weight that connects bulk valence and conduction bands, which we attribute largely to the topological ESs located at structural domain boundaries of bismuthene~\cite{stuhler2020interacting}. Tracking the full relaxation pathway of hot photocarriers across the entire first Brillouin zone (BZ) reveals a fast depletion of the transient conduction band population within $\sim1$~ps, as the in-gap states enable a highly efficient relaxation of excited carriers -- incompatible with the slow recombination observed in topologically trivial indirect semiconductors.

Bismuthene is epitaxially grown on SiC(0001) substrates (see methods). High-quality sample surfaces with low defect rates are confirmed using scanning tunneling microscopy (STM) and low-energy electron diffraction (LEED), see Supplementary Fig.~S1.

\begin{figure*}[!tbh]
\centering
\includegraphics[width=1\textwidth]{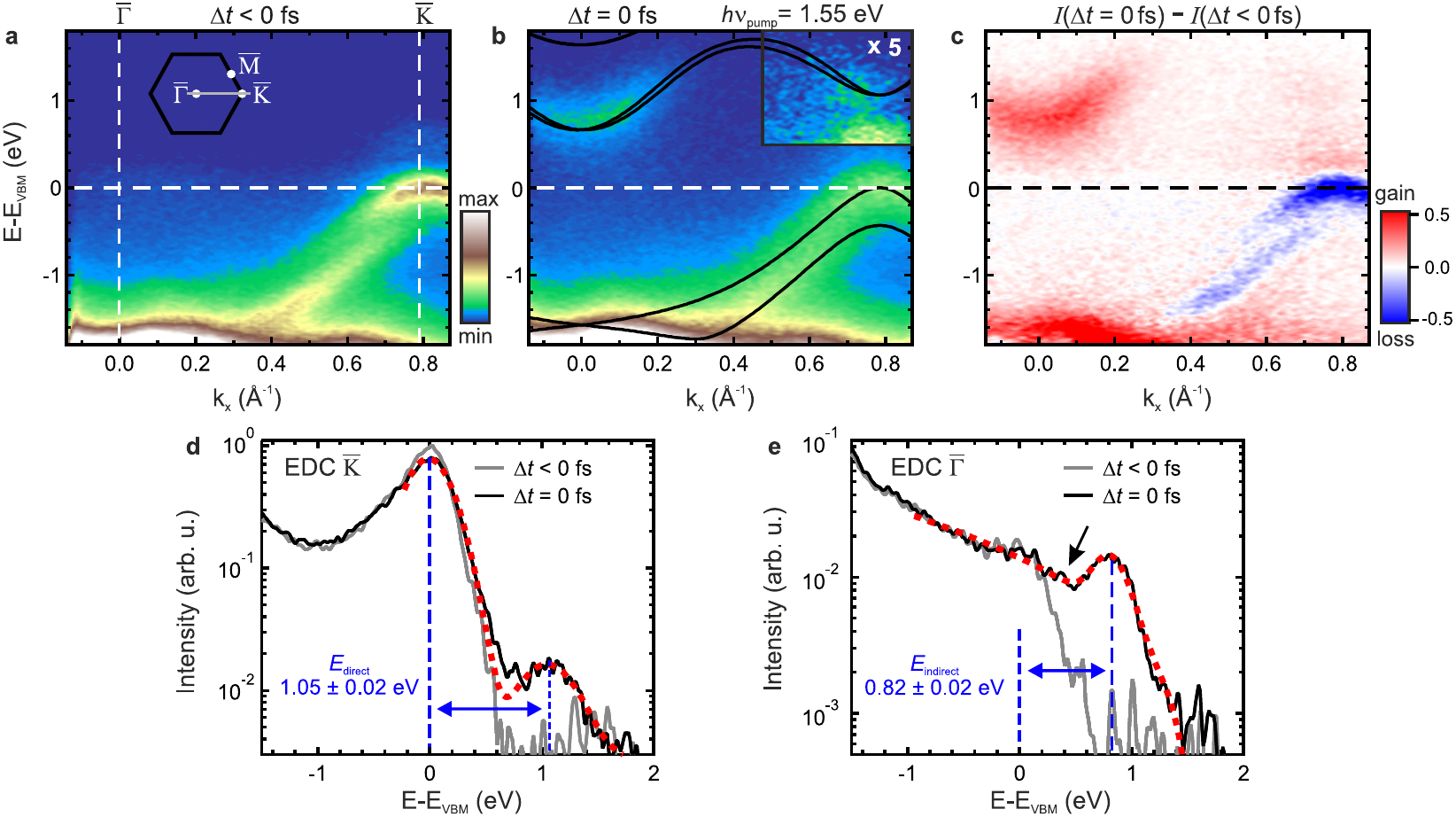}
\caption{Electronic band-structure maps. (a) False-color plots of the trARPES measurements of bismuthene along the $\overline{\Gamma}$-$\overline{\mathrm{K}}$ direction (grey line in inset) in equilibrium and (b) after optical excitation ($h\nu=1.55$\,eV, incident fluence $F$=0.50\,mJ\,cm$^{-2}$, $\Delta t=-40\ldots+40$\,fs). The intensity in the inset is enhanced by a factor of 5. DFT band structure calculations (black) are adopted from Reis et al.~\cite{Reis2017BismutheneOA}. (c) Differential photoemission intensity (pre-excitation signal subtracted) at $\Delta t = 0$\,fs. (d) Energy distribution curves (EDCs) at $\overline{\mathrm{K}}$ and (e) at $\overline{\Gamma}$ in equilibrium and after weak excitation ($F$=0.14\,mJ\,cm$^{-2}$, $\Delta t=-40\ldots+40$\,fs, momentum-integration $\pm0.05$\,\AA$^{-1}$). The red dashed curves mark best fits as described in the text. The black arrow indicates the in-gap intensity extending into the conduction band upon photoexcitation (see discussion). The direct and indirect band gaps are marked in blue.}
\label{fig:data_Phoibos}
\end{figure*}

We begin by mapping the electronic band structure of bismuthene upon photoexcitation, as shown in Fig.~\ref{fig:overview}b. Strong spin-orbit coupling in combination with covalent bonding of the Bi atoms with the substrate opens a large band gap in the Dirac-like crossing at the $\overline{\mathrm{K}}$ points of the hexagonal BZ~\cite{li2018theoretical}. Excitation with near-infrared optical pulses lifts charge carriers across the bulk band gap and transiently populates conduction band states localized at $\overline{\mathrm{K}}$ and, more pronounced, at the $\overline{\Gamma}$ points of the first and second BZs 1\,eV above the valence-band maximum (VBM). Next, we focus on a momentum cut along the $\overline{\Gamma}$-$\overline{\mathrm{K}}$ direction, which features the region of the direct optical interband transition near $\overline{\mathrm{K}}$ and the conduction-band minimum (CBM) at $\overline{\Gamma}$. Consistent with previous studies~\cite{Reis2017BismutheneOA}, the equilibrium band structure of bismuthene features sharp, spin-orbit split low-energy valence bands at $\overline{\mathrm{K}}$ (Fig.~\ref{fig:data_Phoibos}a). Upon optical excitation, a weak excited carrier population at $\overline{\mathrm{K}}$ and a distinct dispersive band at the CBM at $\overline{\Gamma}$ emerge (Figs.~\ref{fig:data_Phoibos}b-c). Concurrently, the valence bands at $\overline{\mathrm{K}}$ are depleted by the optical transition (blue colored region in Fig.~\ref{fig:data_Phoibos}c), and their band width broadens due to scattering of the photoholes with excited quasiparticles~\cite{kanasaki2018ultrafast,dendzik2020corelevel}.

Band-gap renormalization by photo-doping is particularly pronounced in 2D materials due to reduced charge carrier screening~\cite{liu2019direct, Ulstrup2016Jun}. To minimize this effect, we extract the direct and indirect band gaps \red{at temporal pump-probe overlap} using a low incident fluence of 0.14\,mJ\,cm$^{-2}$, as shown in Figs.~\ref{fig:data_Phoibos}d-e. At $\overline{\mathrm{K}}$, we find a direct band gap of 1.05$\pm$0.02\,eV, extracted from the peak positions of Gaussian fits to the upper spin-orbit split band at the VBM and to the lowest-lying CB, which is in excellent agreement with density functional theory (DFT) calculations (1.07\,eV~\cite{Reis2017BismutheneOA}). The indirect band gap between the VBM at $\overline{\mathrm{K}}$ and the CBM at $\overline{\Gamma}$, extracted using a Gaussian fit with an exponentially decaying background, amounts to 0.82$\pm$0.02\,eV, which is in reasonable correspondence with the DFT value of 0.67\,eV~\cite{Reis2017BismutheneOA}. Furthermore, the experimental value also perfectly agrees with the momentum-integrated bulk band gap of $\sim0.8$\,eV obtained from STS measurements~\cite{Reis2017BismutheneOA}. \red{We note that already at a low fluence of 0.14\,mJ\,cm$^{-2}$, the indirect quasiparticle band gap is weakly renormalized by few 10~meV within 100~fs, resulting from the increased screening by quasi-free photocarriers (Supplementary Fig.~S2). For moderate fluences, we observe a significant initial band-gap reduction by $\sim150$~meV, followed by a transient recovery, in agreement with previous studies of 2D semiconductors~\cite{liu2019direct, Ulstrup2016Jun}.} Additional measurements using 3.1\,eV optical excitation, providing a larger view into the dispersion of the lowest-energy conduction band, are shown in Supplementary Fig.~S3.

\begin{figure*}[!t]
\centering
\includegraphics[width=1\textwidth]{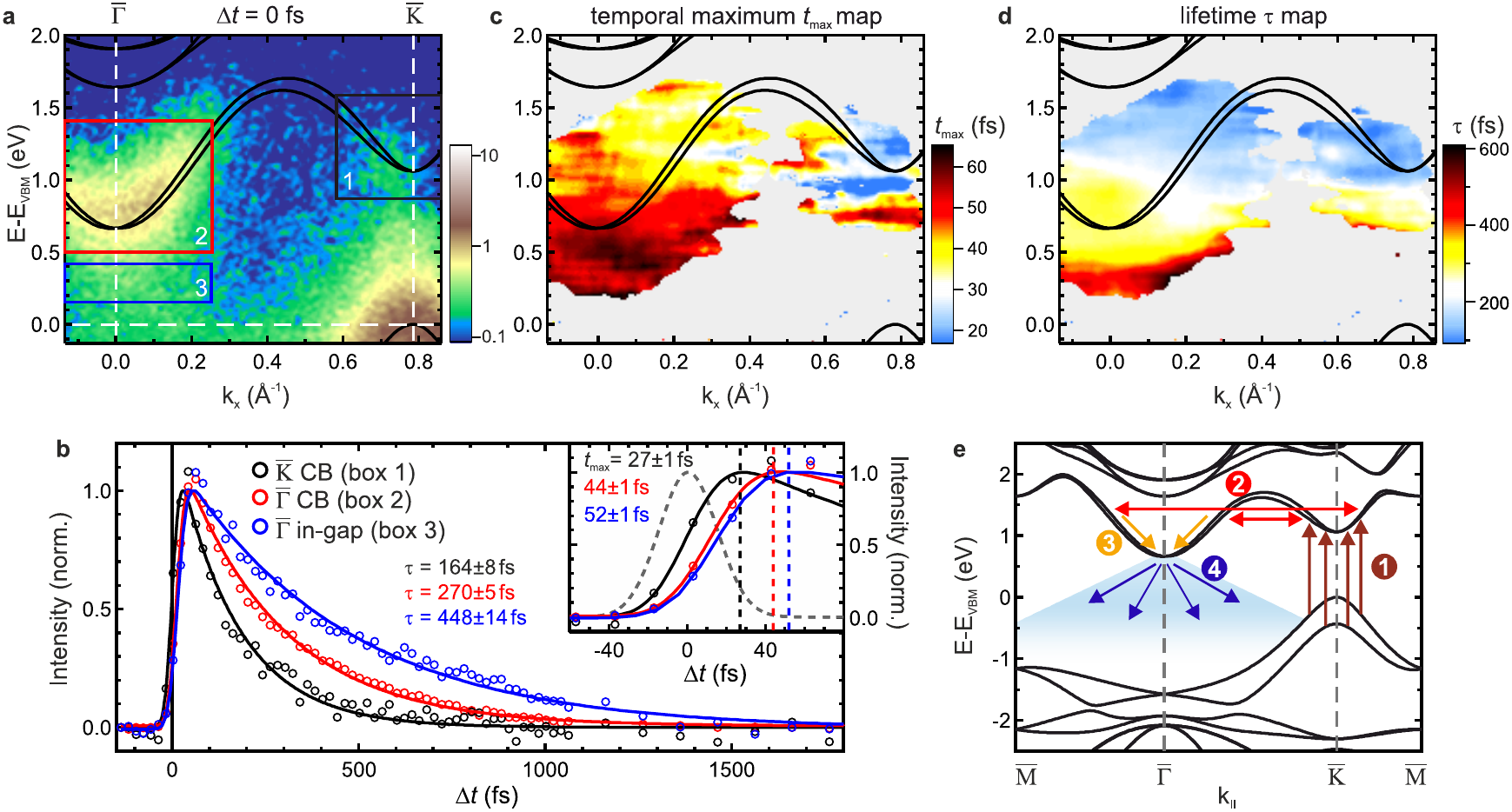}
\caption{Carrier relaxation dynamics. (a) Excited-state band dispersion after 1.55\,eV optical excitation ($F$=0.50\,mJ\,cm$^{-2}$). (b) Normalized photoemission intensities corresponding to boxes 1 to 3 indicated in panel (a) as a function of pump-probe delay. The solid lines show best fits using a single-exponential decay convolved with a Gaussian (Gaussian width as free parameter). The fit parameters $t_{\mathrm{max}}$ (temporal intensity maximum) and $\tau$ ($1/e$ decay constant) are given with one standard deviation as uncertainty. Inset: dynamics near $\Delta t=0$\,fs. The grey dashed line indicates the temporal profile of the pump-laser pulse. (c) Temporal maximum $t_{\mathrm{max}}$ and (d) carrier lifetimes $\tau$ from bin-wise energy- and momentum-dependent decay fits. For this, the transient photoemission intensities are extracted across the energy-momentum region shown in panel (a) using a sliding-window integration \red{($\Delta E = 0.1$\,eV, $\Delta k_\mathrm{x}=0.15$\,\AA$^{-1}$)} and fitted using the function described above. Regions with \red{low photoemission intensity or} large fit uncertainties \red{($\sigma_{t_{\mathrm{max}}}>10$\,fs, $\sigma_\tau>40$\,fs)} are masked in grey.  (e) Schematic scattering processes within the DFT band structure (see text). The in-gap states are indicated in blue.}
\label{fig:relaxation}
\end{figure*}

Next, to elucidate the quasiparticle scattering channels in the photocarrier relaxation processes of bismuthene, we investigate the excited-state population dynamics after 1.55\,eV optical excitation along the $\overline{\Gamma}$-$\overline{\mathrm{K}}$ direction (Fig.~\ref{fig:relaxation}a). As the transient photoemission intensities in Fig.~\ref{fig:relaxation}b show, a CB population builds up first near $\overline{\mathrm{K}}$ (box 1 in Fig.~\ref{fig:relaxation}a), reaching its maximum intensity at $27$\,fs. The apparent delay with respect to temporal pump-probe overlap is due to a build-up of the excited-state population at the $\overline{\mathrm{K}}$ valley until the end of the pump laser pulse, see the inset in Fig.~\ref{fig:relaxation}b. Subsequently, carriers appear at $\overline{\Gamma}$ (box 2) with a delay of few 10\,fs, and lastly a faint intensity within the bulk band gap at $\overline{\Gamma}$ (box 3) builds up, followed by a complete recovery on a timescale of $\sim1$\,ps. We quantify these dynamics by employing single-exponential decay fits, characterized by the time delay at which the excited-state population reaches the maximum, $t_\mathrm{max}$, and the $1/e$ lifetime $\tau$. To establish the full energy- and momentum-dependent scattering pathway, we extend this evaluation of three exemplary areas by fitting the transient intensity of each energy-momentum bin across Fig.~\ref{fig:relaxation}a using a sliding integration window. The resulting energy-momentum maps of the fit parameters $t_\mathrm{max}$ and $\tau$ allow us to track the arrival time of excited carriers in energy-momentum space (Fig.~\ref{fig:relaxation}c) and provide a concise overview of the lifetimes associated with particular states (Fig.~\ref{fig:relaxation}d). 

Combining the results of both maps yields a detailed picture of the complete carrier relaxation pathway, schematically depicted in Fig.~\ref{fig:relaxation}e: (1) Carriers are initially injected by a vertical interband transition into the CB near $\overline{\mathrm{K}}$ using 1.55\,eV radiation.
(2) The hot electrons redistribute by intervalley scattering, which spreads the carriers over an extended momentum region into the $\overline{\Gamma}$ valley on a 10 fs timescale, a phenomenon commonly observed in photoexcited semiconductors~\cite{bertoni2016, hein2016momentum, madeo2020, dong2020measurement}. (3) Subsequently, hot carriers relax toward the CBM at $\overline{\Gamma}$ via electron-electron and electron-phonon scattering within $\sim50$\,fs. Although bulk bismuthene exhibits an indirect band gap of nearly 1~eV, the lifetime of the conduction band population is only on the order of few 100\,fs -- an orders of magnitude faster relaxation than in conventional indirect semiconductors~\cite{schroder2015semiconductor, bertoni2016, wang2015surface, Li2021_PRMat, Lee2021Sep}. (4) These ultrashort lifetimes indicate a highly efficient carrier relaxation. The question naturally arises, which states other than the insulating 2D bulk states in bismuthene could mediate the observed fast decay. Intriguingly, we observe faint gap-filling spectral weight reaching up to the CBM for several 100~fs after photoexcitation (Supplementary Fig.~S4), which we discuss below. A relaxation of the conduction band population through these in-gap states is supported by the fact that they are populated last and feature the longest lifetimes. Finally, within $\sim1.5$\,ps, also the in-gap states above the VBM are fully depleted. Note that the extracted population lifetimes are distinct from single-particle lifetimes that are directly encoded in the electron self-energy, and thus only represent an upper limit for the timescale of scattering processes~\cite{Yang2015, Kemper2018_PRX}.

\begin{figure*}[!t]
\centering
\includegraphics[width=1\textwidth]{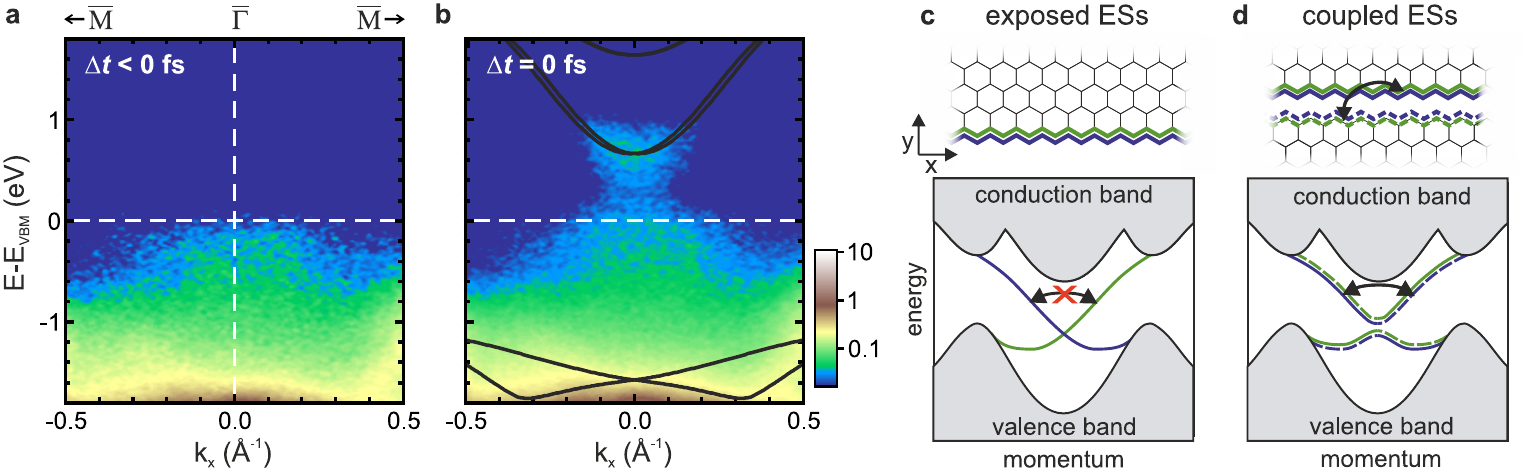}
\caption{In-gap intensity. (a) Photoemission spectra at $\overline{\Gamma}$ in equilibrium and (b) after optical excitation ($h\nu$=1.55\,eV, $F$=0.32\,mJ\,cm$^{-2}$). Faint spectral weight is located in between the band gap predicted by bulk DFT calculations (black), connecting the valence and conduction states. Visibility of the in-gap states is enhanced by a logarithmic color scale. (c and d) Top: Sketch of helical ESs (blue, green) at an exposed sample edge with zigzag termination and of coupled ESs at a domain boundary, respectively. Bottom: Schematic dispersion of infinitely extended exposed and coupled 1D ESs, respectively. The projected bulk band structure is indicated in grey. The hybridization of coupled ES pairs opens an energy gap and lifts spin-momentum locking, enabling single-particle backscattering, indicated by the black arrow. The size of the gap opening at the crossing of the ES dispersion, however, is expected to be significantly lower than our experimental energy resolution of $\sim150$\,meV. Adopted from Refs.~\cite{Reis2017BismutheneOA} and~\cite{stuhler2020interacting}.}
\label{fig:data_Gamma}
\end{figure*}

Lastly, we examine the in-gap spectral weight and discuss its origin. We find that already in equilibrium a faint intensity is located at $\overline{\Gamma}$ reaching up to $E_\mathrm{VBM}$  (Fig.~\ref{fig:data_Gamma}a). Upon optical excitation, the in-gap intensity extends into the CBM (Fig.~\ref{fig:data_Gamma}b), resulting in the absence of an explicit gap between bulk conduction and valence bands (black arrow in Fig.~\ref{fig:data_Phoibos}e).
While it may seem obvious to assign the in-gap feature with the topological ESs that were directly probed in previous STM/STS studies~\cite{Reis2017BismutheneOA, stuhler2020tomonaga, stuhler2020interacting}, a careful examination is required.

For that, we compare the relaxation dynamics and the in-gap feature for bismuthene grown on two different substrates types, i.e., intentionally miscut and planar SiC substrates (Supplementary Fig.~S1). Using a miscut substrate yields a high density of unidirectional, topologically protected exposed ESs along the parallel substrate step edges, as illustrated in Fig.~\ref{fig:data_Gamma}c, while bismuthene prepared on a planar substrate features only a negligible density of (randomly oriented) substrate steps with exposed ESs. Yet, for both substrate types, additional topological ESs arise at structural phase-slip domain boundaries within the 2D bulk of bismuthene. These domain boundaries result from the fact that the $(\sqrt{3}\times\sqrt{3})$ Bi honeycombs can have three distinct registries with respect to the substrate lattice, causing the growth of domains. Pairs of helical ESs emerge at the zigzag edges on either side of these boundaries, see Fig.~\ref{fig:data_Gamma}d, lifting the topological protection through mutual hybridization and leading to a mixing of different helicities, as confirmed by local tunneling spectroscopy~\cite{stuhler2020interacting}.

The presented characterization of the in-gap feature and relaxation dynamics was conducted on bismuthene prepared on a miscut substrate. However, in bismuthene prepared on a planar substrate featuring only a negligible density of exposed ESs, equivalent relaxation dynamics (Supplementary Fig.~S5) and a similar in-gap feature (Supplementary Figs.~S4 and S6) are observed. This leads us to the conclusion, that the in-gap intensity primarily originates from topological ESs that arise at structural domain boundaries, which are present in both planar and miscut samples. The faint intensity of the in-gap states on the order of a few percent of the bismuthene bulk bands at $\overline{\mathrm{K}}$ is consistent with the assignment to coupled ESs, as the domain boundaries constitute only a fraction of the probed surface. We further conclude that the relaxation of excited-state charge carriers must be strongly facilitated by the quasi-metallic density of states observed at such domain boundaries, enabling the rapid depletion of the conduction band population, analogous to the depletion of conduction band populations by topological surface states in photoexcited 3D TIs~\cite{sobota2012ultrafast}. Since we do not observe large deviations for the decay times in the case of the bismuthene sample grown on a miscut substrate (Supplementary Fig.~S5), any definite conclusion on the additional role of the exposed topological ESs is difficult to draw at this point. Future developments in (time-resolved) nano-ARPES featuring a nanometer spatial resolution may allow isolating the spectral features of exposed ESs at substrate terrace steps and individual pairs of coupled ESs at domain boundaries.

While the coupled ESs at domain boundaries consistently explain the in-gap spectral weight and short photocarrier lifetimes, several alternative scenarios can potentially induce a continuous in-gap intensity. First, impurities and defects may provide additional states within the bulk band gap. However, STM/STS studies show that only domain boundaries and exposed sample edges host in-gap states, while all other sample regions remain fully gapped~\cite{Reis2017BismutheneOA, stuhler2020tomonaga, stuhler2020interacting}. In addition, the observed confinement of the in-gap intensity to the momentum-region near $\overline{\Gamma}$ speaks against impurities as primary origin, as such defect states typically lack a clear momentum dependence~\cite{Strocov2019impuritybands}.
Second, a large energy line-width of the bismuthene bulk valence band may lead to a metal-like extension of intensity up to the Fermi level. However, analogous to the first line of argument, spatially-resolved STS measurements unambiguously demonstrate an insulating behavior in bulk regions~\cite{Reis2017BismutheneOA}, indicating that the observed continuous in-gap intensity is not connected to bulk bismuthene, but rather to ESs. Thus, by excluding alternative interpretations and consistent with earlier local tunneling spectroscopy results~\cite{stuhler2020interacting}, we assign the in-gap spectral weight largely to coupled ESs at domain boundaries. \red{As ESs forming along extended 1D defects critically limit the lifetime of excited carriers and may also pose challenges for applications utilizing the spin-selective transport along exposed sample edges~\cite{pezo2021disorder, tiwari2019carrier}, our study underlines the need for high-quality sample surfaces.}

In conclusion, we experimentally map out the electronic band structure of the quantum spin Hall insulator bismuthene after near-infrared photoexcitation and determine the direct and indirect band gaps to $\sim$1.1\,eV and $\sim$0.8\,eV, respectively. Analysis of the microscopic scattering pathway of hot photocarriers reveals exceptionally fast carrier relaxation dynamics governed by faint in-gap states located within the indirect band gap, which are in correspondence with the topological edge states arising at bismuthene domain boundaries. The demonstration of a large fundamental band gap and the in-gap spectral weight persisting at room temperature and under strong optical excitation highlights the promising role of bismuthene as an ambient-condition quantum spin Hall candidate. \red{Additionally, due to the exceptionally large band gap, bismuthene serves as a unique platform for optically addressing novel functionalities based on the topological edge states and for studying excitons in a topologically nontrivial system. Our insights gained on quasiparticle scattering processes lay the basis for future studies of sub-band-gap excitations and optical control schemes of edge-state currents~\cite{McIver2012, Kuroda2017_PRB}}.

\section*{Methods}
\subsection*{Sample preparation and STM measurements.} 
Bismuthene was epitaxially grown on n-doped 4H-SiC(0001) substrates (0.01 - 0.03 $\Omega\cdot$cm, carrier concentration $\sim10^{18} - 10^{19}$ cm$^{-3}$, planar and 4° miscut) in ultra-high vacuum $<10^{-10}$\,mbar. Prior to growth, a smooth H-terminated SiC surface was prepared by hydrogen-based dry-etching. Growth was performed at $\sim$600\,°C to thermally desorb the surface H-termination, while simultaneously offering Bi atoms from a commercial effusion cell~\cite{Reis2017BismutheneOA}. Successful growth of low-defect bismuthene samples was verified using low-energy electron diffraction and scanning tunneling microscopy.

\subsection*{Time-resolved ARPES measurements.} 
After characterization, the samples were transferred to the trARPES setup using a UHV suitcase at  $p<10^{-10}$\,mbar. All measurements were performed at room temperature using a laser-based high-harmonic-generation trARPES setup (p-polarized probe at $h\nu_{\mathrm{probe}}$=21.7\,eV, s-polarized pump at $h\nu_{\mathrm{pump}}$=1.55 / 3.10\,eV, 500\,kHz repetition\@ rate, $\Delta E\sim$150\,meV, $\Delta t\sim$40\,fs) with a 6-axis manipulator (SPECS Carving)~\cite{puppin2019}. Photoelectrons were detected either with a hemispherical analyzer (SPECS Phoibos 150) or a time-of-flight momentum microscope (SPECS METIS 1000)~\cite{maklar2020}. The momentum microscope allows for a parallel acquisition of the 3D photoelectron distribution $I(E_\mathrm{kin}, k_\mathrm{x}, k_\mathrm{y})$ across a large energy and momentum range, and was thus utilized for overview measurements of the electronic band structure (Fig.~\ref{fig:overview}). In contrast, the hemispherical analyzer allows for fast data acquisition within a limited energy-momentum window, and was thus used to map selected high-symmetry directions (Figs.~\ref{fig:data_Phoibos}-\ref{fig:data_Gamma}). The data presented in Fig.~\ref{fig:overview} were acquired on a planar substrate; Figs.~\ref{fig:data_Phoibos}-\ref{fig:data_Gamma} on a miscut substrate. The XUV probe spot size (FWHM) was $\sim80 \times 80$\,\textmu m$^2$. The pump spot sizes were $\sim260  \times 200$\,\textmu m$^2$ ($h\nu$=1.55\,eV) and $\sim510  \times 475$\,\textmu m$^2$ ($h\nu$=3.10\,eV). All fluences stated in the text correspond to incident fluences. Temporal pump-probe overlap was determined from the pump-laser-induced depletion of the valence band population, as shown in Supplementary Fig.~S7.

\section*{Acknowledgment}
This work was funded by the Max Planck Society, the European Research Council (ERC) under the European Union's Horizon 2020 research and innovation program (Grant No. ERC-2015-CoG-682843 and OPTOlogic 899794) and the German Research Foundation (DFG) under the Emmy Noether program (Grant No. RE 3977/1), the SFB/TRR 227 "Ultrafast Spin Dynamics" (Project ID 328545488, projects A09 and B07), the DFG research unit FOR 1700, the Priority Program SPP 2244 (project 443366970), the W\"urzburg-Dresden Cluster of Excellence on Complexity and Topology in Quantum Matter ct.qmat (EXC 2147, Project ID 390858490) as well as through the Collaborative Research Center SFB 1170 ToCoTronics (Project ID 258499086). S.B. acknowledges financial support from the NSERC-Banting Postdoctoral Fellowships Program. M.D. acknowledges financial support from the Göran Gustafsson Foundation. T.P. acknowledges financial support from the Alexander von Humboldt Foundation.

\section*{Data availability}
\red{All data that support the findings of this study are publicly available at the Zenodo data repository, https://doi.org/10.5281/zenodo.5512069.}

\section*{Author contributions}
J.M., M.D., R.S., S.D., S.B., T.P., A.N. and L.R. carried out the trARPES experiments; J.M. analyzed the trARPES data with support from R.S.;  R.S. prepared and characterized the samples and analyzed the STM data; G.L. performed the DFT calculations; J.M. wrote the manuscript with support from L.R., R.S., R.E. and R.C.; L.R., R.E., M.W. and R.C. provided the experimental infrastructure; all authors commented on the paper.

\section*{Competing Interests}
The authors declare that they have no competing financial interests.

\printbibliography

\include{supplement}
\end{document}

%% file: supplement.tex

\clearpage
\beginsupplement
\appendix
\section*{Supporting Information}
\subsection*{Supplementary Figures}
\label{supp:supp_figures}

\begin{figure*}[!ht]
\centering
\includegraphics[scale=0.95]{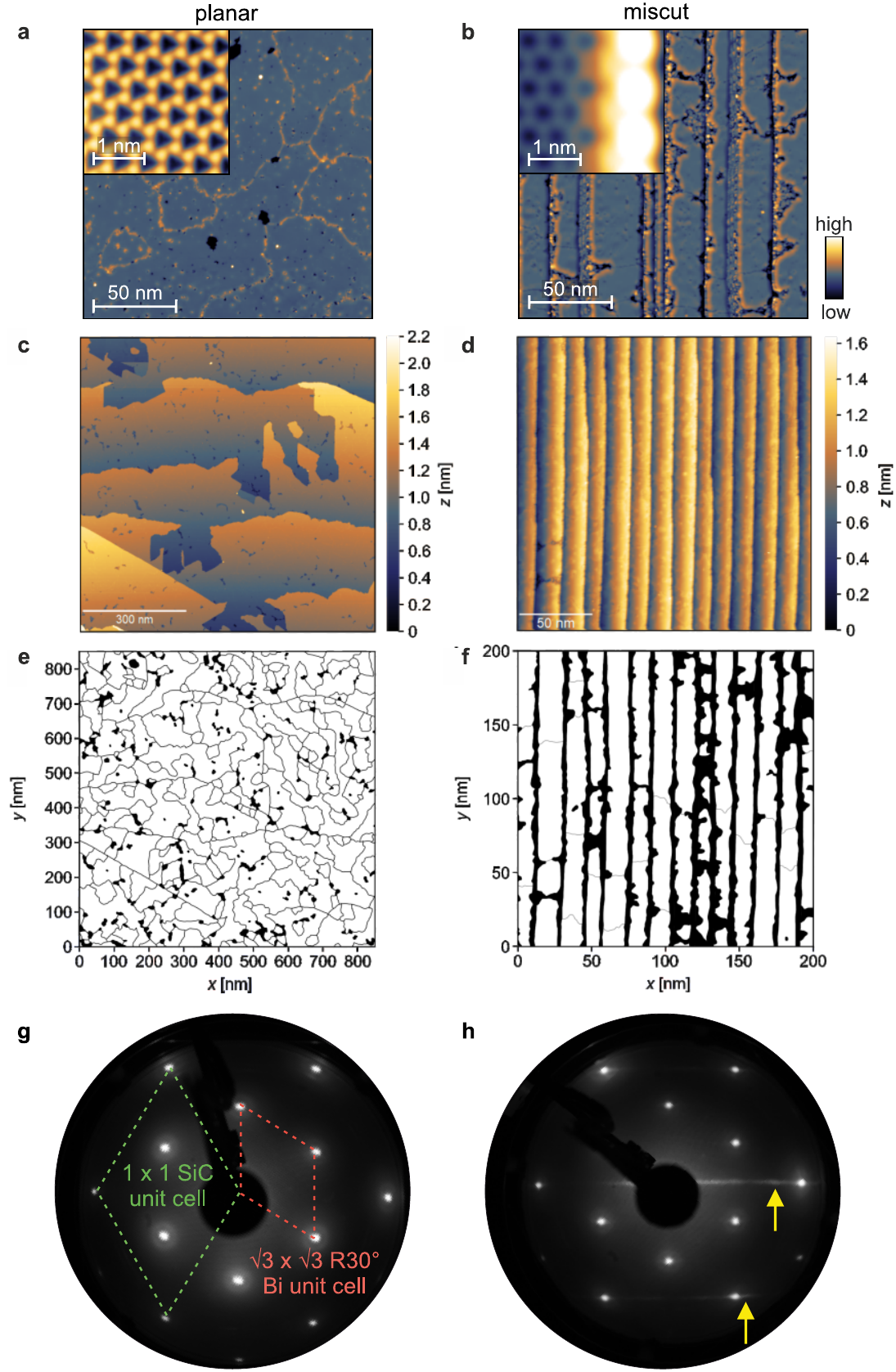}
\caption{Surface characterization of bismuthene samples. (a) STM constant-current image of bismuthene on a planar SiC substrate. A meandering network of domain boundary segments (bright ridges) intersects the bismuthene film into connected domains. Inset: Close-up of the Bi honeycomb lattice. (b) STM image of bismuthene on a 4° miscut substrate featuring unidirectional SiC terrace steps every $\sim15$\,nm that induce exposed bismuthene edges. Inset: Bi honeycombs near a step edge. While both the morphology of the planar and the miscut bismuthene samples feature domain boundaries, the miscut bismuthene sample exhibits a significantly larger exposed edge density. Scan parameters: $V_\text{set} = 3.0$\,V, $I_\text{set} = 50$\,pA, $T=4.35$\,K; insets: $V_\text{set}=-0.8$\,V,  $I_\text{set}=100$\,pA, $T=4.35$\,K. (c and d) Overview STM constant current images of bismuthene on a planar and a miscut substrate, respectively. Scan parameters: (c) $V_\text{set} = 3.0$\,V, $I_\text{set} = 50$\,pA, $T = 4.35$\,K; (d) $V_\text{set} = 2.6$\,V, $I_\text{set} = 30$\,pA, $T = 4.35$\,K. (e and f) Binary masks marking domain boundaries and defective areas on the planar and miscut samples from (c) and (d), respectively. (g) Low-energy electron diffraction of bismuthene on a planar SiC substrate recorded at an energy of 48\,eV and (h) on a miscut substrate at 50\,eV. Sharp, intense diffraction spots and a weak diffuse background signal indicate high-quality sample surfaces. The yellow arrows in (h) mark stripe-like elongations of the SiC spots corresponding to unidirectional substrate step edges. Exemplary SiC 1x1 and Bi $\sqrt{3}$x$\sqrt{3}$ R30° reciprocal unit cells are indicated.}
\label{fig:S_LEED}
\end{figure*}

\begin{figure*}[!ht]
\centering
\includegraphics[]{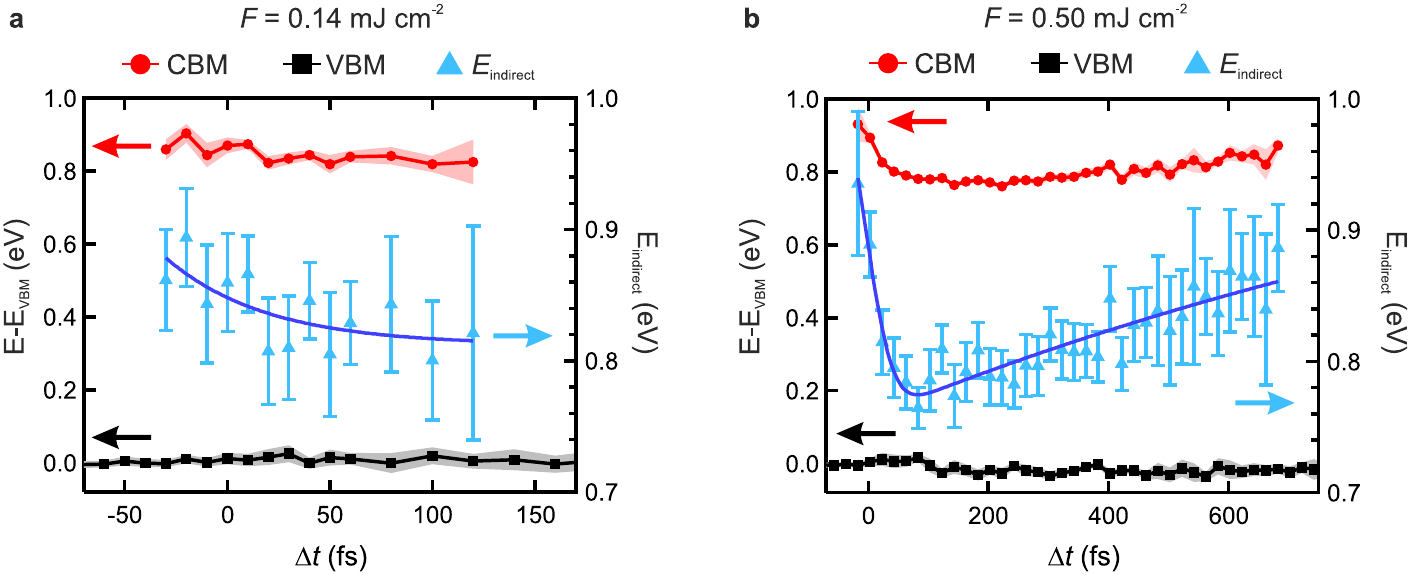}
\caption{\red{Renormalization of the quasiparticle band gap. Position of the CBM and VBM (left axis) and extracted indirect band gap (CBM-VBM difference, right axis) as a function of pump-probe delay for (a) low and (b) moderate incident fluence. When applying a low fluence of $F=0.14$\,mJ\,cm$^{-2}$ (panel a), the VBM position remains approximately constant while the CBM position shifts slightly downwards within 100~fs, resulting in a reduction of the indirect band gap by $\sim40$~meV. However, in this fluence regime, the limited number of hot carriers in combination with their fast relaxation allows for reliable tracking of the CBM only up to $\sim120$~fs. For a moderate fluence of $F=0.50$\,mJ\,cm$^{-2}$ (panel b), we observe a pronounced time-dependent band-gap renormalization, as the CBM undergoes a significant shift while also the position of the VBM changes with $\Delta t$. The photoexcited quasi-free carriers initially increase the screening of the Coulomb interactions, transiently reducing the effective band-gap size by $\sim150$~meV. As the system relaxes to equilibrium, the band gap recovers with increasing $\Delta t$. The solid blue lines in (a) and (b) serve as guides to the eye. The band positions were extracted using Gaussian fits as discussed in the main text. The error bands and bars correspond to one standard deviation resulting from the Gaussian fits.}}
\label{fig:S_renormalization}
\end{figure*}

\begin{figure*}[!ht]
\centering
\includegraphics[]{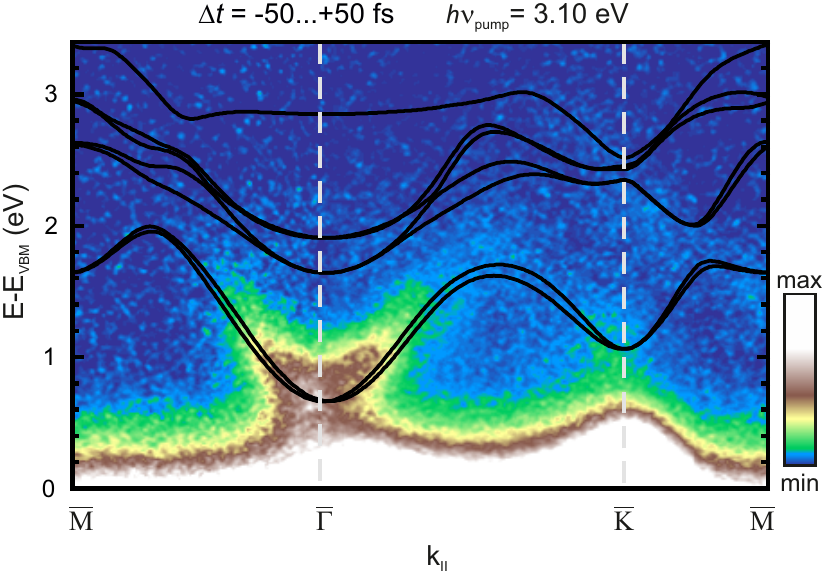}
\caption{Conduction band dispersion after photoexcitation. Photoemission intensity along the high-symmetry momentum directions after 3.1\,eV optical excitation ($F$=0.03\,mJ\,cm$^{-2}$) at temporal pump-probe overlap. DFT calculations are shown in black.}
\label{fig:S_3eV}
\end{figure*}

\begin{figure*}[!ht]
\centering
\includegraphics[width=\textwidth]{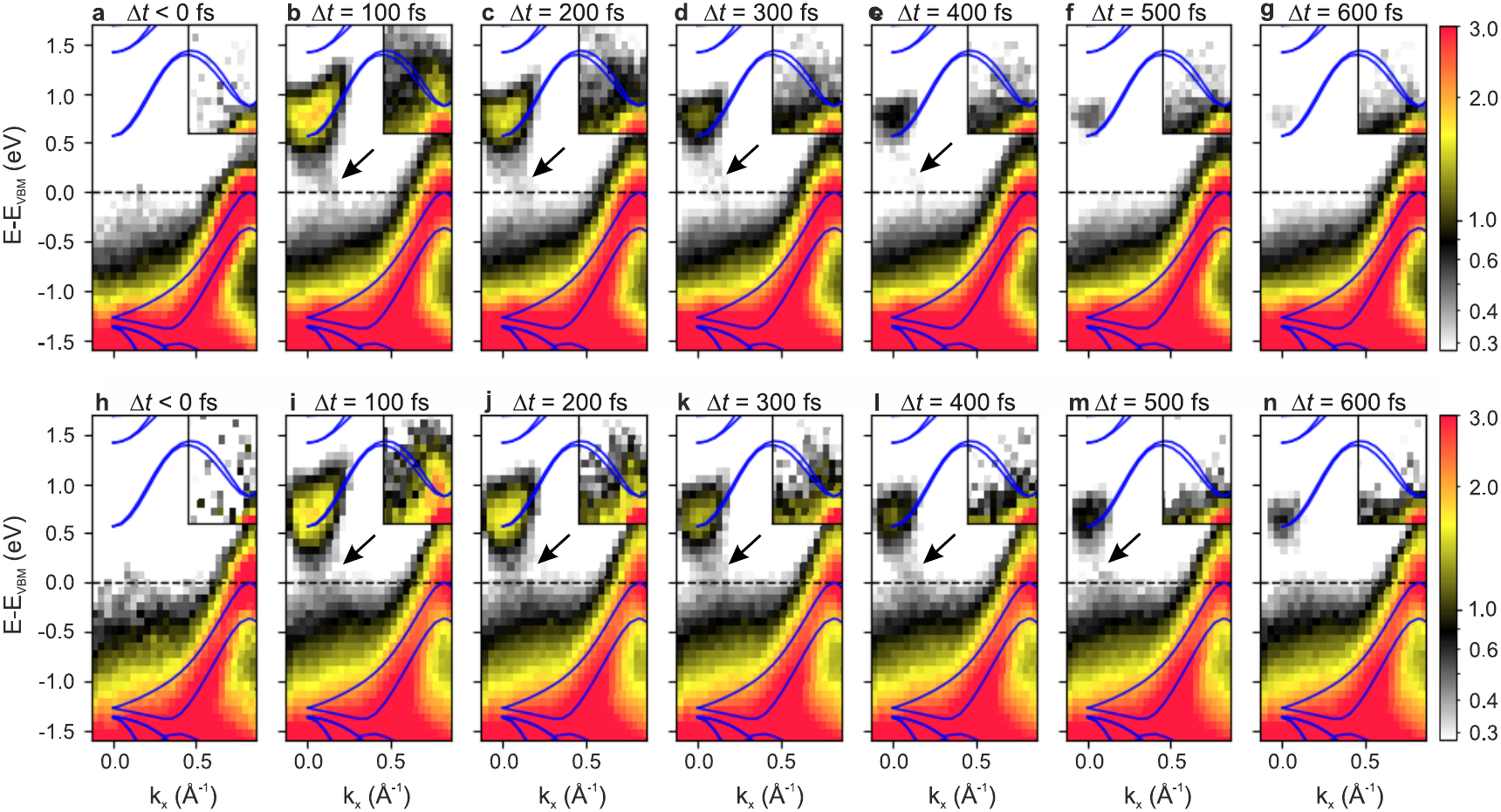}
\caption{Dynamic band-structure maps. (a-g) False-colour plots of the trARPES measurements of bismuthene on a miscut substrate along the $\overline{\Gamma}$-$\overline{\mathrm{K}}$ direction for selected time delays (time integration window of 200~fs, $h\nu=1.55$\,eV, $F$=0.50\,mJ\,cm$^{-2}$).  (h-n) Equivalent measurements for bismuthene on a planar substrate. The black arrows indicate the in-gap intensity near $\overline{\Gamma}$. The intensity in the insets at $\overline{\mathrm{K}}$ is enhanced. DFT band structure calculations are shown in blue.}
\label{fig:S_dynamics}
\end{figure*}

\begin{figure*}[!ht]
\centering
\includegraphics[]{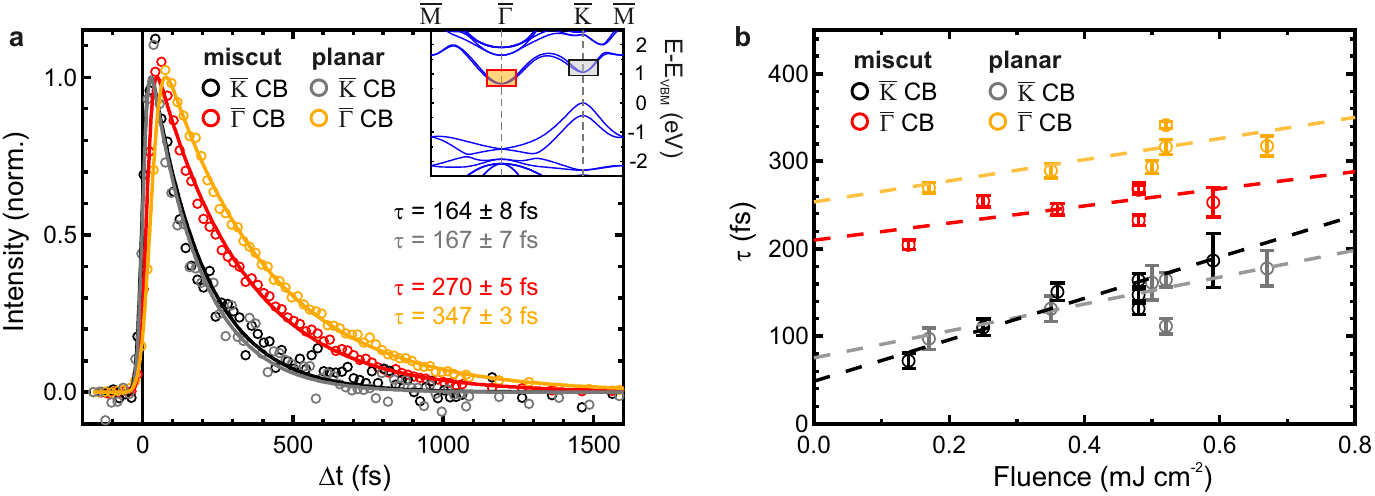}
\caption{Photocarrier lifetimes. (a) Normalized photoemission intensities of the conduction band populations at $\overline{\mathrm{K}}$ and $\overline{\Gamma}$ for bismuthene on miscut and planar substrates ($F$=0.50\,mJ\,cm$^{-2}$) versus pump-probe delay. The inset indicates the energy-momentum regions of interest within the DFT band structure of the respective time traces (equivalent to boxes 1 and 2 in Fig.~3a). Single-exponential decay fits reveal similar 1/$e$ lifetimes of the populations at $\overline{\mathrm{K}}$ for both substrate types, and slightly increased lifetimes at $\overline{\Gamma}$ for the planar substrate. (b) Extracted lifetimes for both substrate types as function of incident fluence. The dashed lines serve as guides to the eye. While the overall lifetime increases with fluence, the lifetime at $\overline{\Gamma}$ for the planar substrate is systematically higher with respect to the miscut substrate for all applied fluences. The error bars correspond to one standard deviation of the fit parameter $\tau$.}
\label{fig:S_lifetimes_comparison}
\end{figure*}

\clearpage
\begin{figure*}[!ht]
\centering
\includegraphics[]{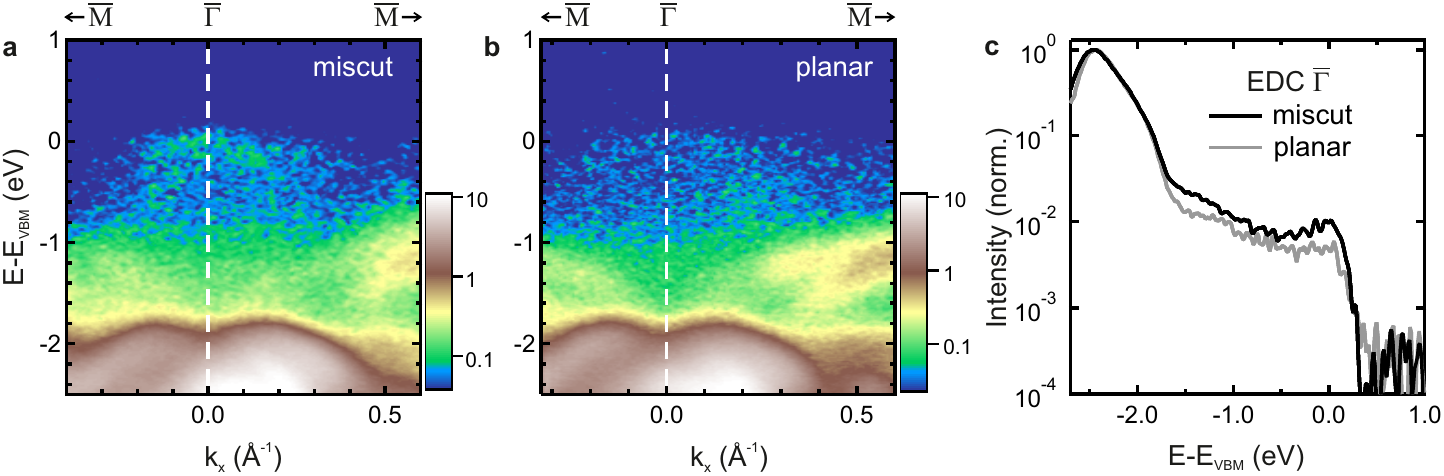}
\caption{In-gap feature at $\overline{\Gamma}$. (a) Equilibrium band dispersion of bismuthene on a miscut substrate. A faint feature near the VBM is identified, which we assign to topological ESs.
(b) Band dispersion of bismuthene on a planar substrate, exhibiting a similar, slightly less pronounced feature at $\overline{\Gamma}$. (c) Corresponding EDCs at $\overline{\Gamma}$. Both samples show a faint but distinct intensity up to $E_\mathrm{VBM}$, which is roughly a factor 2 more intense for the miscut substrate.}
\label{fig:S_Gamma_feature}
\end{figure*}

\subsection*{Determination of temporal pump-probe overlap}
To determine the temporal pump-probe overlap $\Delta t=0$\,fs, we extract the initial depletion of the valence band population at $\overline{\mathrm{K}}$ (red box in Supplementary Fig.~\ref{fig:S_t0}a) resulting from the vertical optical transition. The extracted photoemission intensity as function of pump-probe delay is fitted using an error-function, see Supplementary Fig.~\ref{fig:S_t0}b. Here, the central position of the error function corresponds to the temporal peak of the optical pump pulse, $t_0$. For all measurements presented in this manuscript, we calibrate the pump-probe delay such that $\Delta t=0$\,fs corresponds to $t_0$.

\begin{figure*}[!ht]
\centering
\includegraphics[]{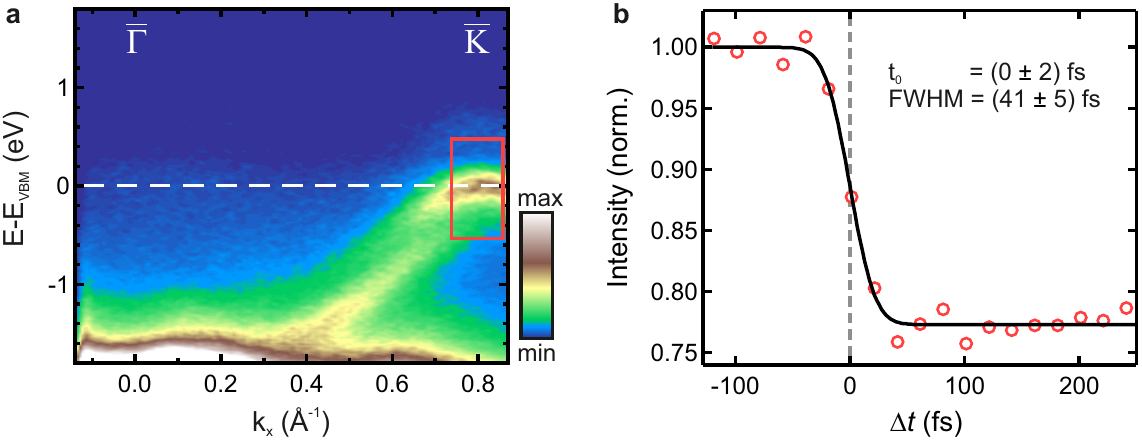}
\caption{Arrival time of the optical excitation pulse. (a) Equilibrium ARPES measurement of the $\overline{\Gamma}$-$\overline{\mathrm{K}}$ direction.  (b) Normalized photoemission intensity extracted from the red box in panel (a) as function of pump-probe delay (red circles, $h\nu_\mathrm{pump}=1.55$\,eV, incident fluence $F$=0.50\,mJ\,cm$^{-2}$). The black solid line marks the best fit using an error function. Fit coefficient values of the temporal peak $t_0$ and full width at half maximum of the pump pulse are stated in the figure, respectively. One standard deviation is given as uncertainty.}
\label{fig:S_t0}
\end{figure*}